# Detecting mid-infrared light by molecular frequency upconversion with dual-wavelength hybrid nanoantennas

Angelos Xomalis [1], Xuezhi Zheng [1,2], Rohit Chikkaraddy [1], Zsuzsanna Koczor-Benda [3], Ermanno Miele [1,4,5], Edina Rosta [3], Guy A E Vandenbosch [2], Alejandro Martínez [6], Jeremy J Baumberg [1*]

[1] NanoPhotonics Centre, Cavendish Laboratory, Department of Physics, University of Cambridge, Cambridge, United Kingdom.
[2] Department of Electrical Engineering (ESAT-TELEMIC), KU Leuven, Leuven, Belgium
[3] Department of Physics and Astronomy, University College London, London, United Kingdom
[4] Department of Chemistry, University of Cambridge, Cambridge, United Kingdom
[5] The Faraday Institution, Harwell Science and Innovation Campus, Oxford, United Kingdom
[6] Nanophotonics Technology Center, Universitat Politècnica de València, Valencia, Spain

*Corresponding author. email: jjb12@cam.ac.uk

**ABSTRACT**
**Coherent interconversion of signals between optical and mechanical domains is enabled by optomechanical interactions. Extreme light-matter coupling produced by confining light to nanoscale mode volumes can then access single mid-infrared (MIR) photon sensitivity. Here we utilise the infrared absorption and Raman activity of molecular vibrations in plasmonic nanocavities to demonstrate frequency upconversion. We convert $\lambda$ ~10 µm incoming light to visible via surface-enhanced Raman scattering (SERS) in doubly-resonant antennas that enhance upconversion by $>10^{10}$. We show >200% amplification of the SERS antiStokes emission when a MIR pump is tuned to a molecular vibrational frequency, obtaining lowest detectable powers ~1 µW/µm² at room temperature. These results have potential for low-cost and large-scale infrared detectors and spectroscopic techniques, and bring single-molecule sensing into the infrared.**

One-Sentence Summary:
**Molecules as Mixers.**
**Detecting weak mid-infrared light currently requires expensive cooled devices. Xomalis *et al* instead convert the light to visible frequencies where it is efficiently measured, by using vibrating molecules inside a cavity that traps the multiple frequencies of light simultaneously. Using self-assembled devices integrated onto silicon wafers, they exploit the nonlinear interaction of mechanical vibrations and oscillations of light. This yields the first successful devices in the recently emerging field of molecular optomechanics.**

Infrared spectroscopy delivers information hard to obtain from other frequency bands, such as atmospheric absorption of molecules (greenhouse gases) or thermally-emitted radiation from earth (meteorological maps or imaging wildfires) (*1-5*). While development of mid-IR (MIR) sources evolves, a bottleneck continues to be in producing low-noise room-temperature detectors (*6*). One proposed scheme is to directly upconvert MIR photons into high-energy visible photons that are efficiently detected, potentially delivering single-photon semiconductor-based detectors (*7-9*). Analogous wavelength conversion from microwave to optical frequencies has utilised expensive fabrication and cryogenic temperatures (*10, 11*), as well as LiNbO$_3$ resonators (*12, 13*). Critically, in order to access the efficiencies required, strongly-enhanced light-matter interactions are paramount. Thus, plasmonic devices and planar resonant metasurfaces which confine light have been of interest for MIR integrated detection and biosensing (*14-16*).

A promising approach for detecting infrared radiation through frequency upconversion is via molecular optomechanical coupling (*17*). Optomechanical interactions allow coherent conversion of signals between optical and mechanical domains (Fig. 1). Nanocavities containing vibrating molecules act as mechanical oscillators, with MIR-absorbing infrared vibrational modes probed by a visible laser through their Raman scattering (Fig. 1B). The required interactions can be boosted by using the tight light localisation inside plasmonic nanocavities <100nm across which yield detectable signals even from single vibrational bonds (*18*). The interaction of light and matter in these sub-nm mode volumes gives extreme optomechanical coupling with single MIR-photon sensitivity in principle, but so far only studied theoretically (*17*). The noise-equivalent power of hybrid nanocavity-molecular detectors is predicted to be hundred-fold lower than commercial uncooled detectors.

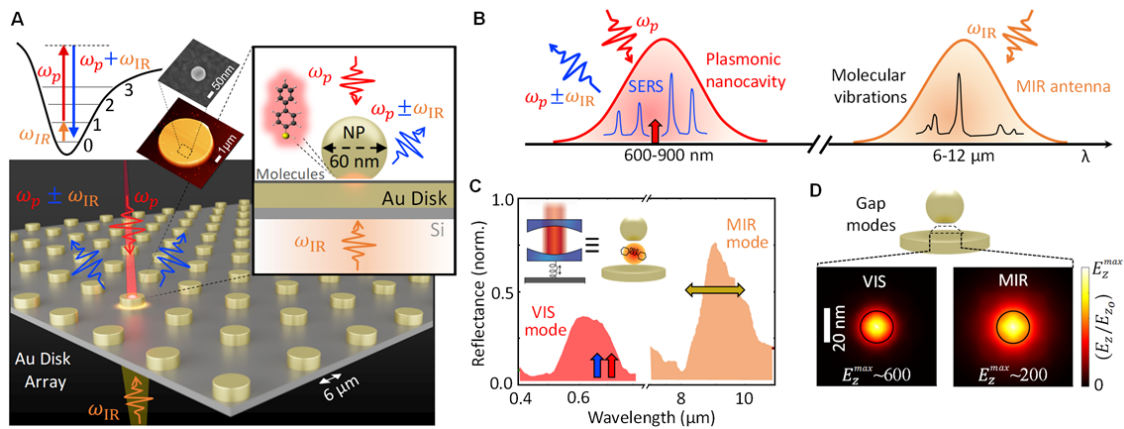

**Fig. 1. Dual-wavelength antenna and frequency upconversion.** (**A**) Pump (MIR) - probe (visible) detection configuration. Inset shows upconversion process, AFM (disk) and SEM (nanoparticle) images, and self-assembled monolayer of biphenyl-4-thiol which creates 1.3nm cavity between 60nm Au nanoparticle and 6µm disk. (**B**) Scheme of MIR to visible up-conversion via molecular optomechanics. (**C**) Experimental reflectance of NPoR resonances at both visible (red) and MIR (orange) wavelengths. Arrows indicate SERS probe wavelength (785nm, red), inelastic scattered light (blue) and MIR tuning range (8.5-12.6µm, yellow). Inset shows equivalence of optomechanical cavity and NPoR. (**D**) Near-field normalised maps of mid-infrared (MIR) and visible gap modes of NPoR. Black circle shows 20 nm nanoparticle facet.

Of vital importance for upconversion efficiency is the optimal spatial overlap of visible and infrared radiation. Plasmonics allows extreme light confinement at visible frequencies, but at longer wavelengths light localisation drops significantly. Achieving light confinement simultaneously in both visible and MIR spectral regions requires a hybrid dual resonator (*19*). Here, this is fulfilled by creating doubly-resonant antennas which focus long- and short-wavelengths into the same active region allowing good optomechanical coupling (Fig. 1D). Their construction combines bottom-up and top-down methods that allow for ease of fabrication and cost-effective large-scale arrays of devices.

To demonstrate MIR detection, we perform surface-enhanced Raman spectroscopy (SERS) on self-assembled molecular monolayers (SAMs) with discrete vibrational absorption modes in the $\lambda$=6-12µm range. Coupling requires matching the optical (infrared absorption) and mechanical (molecular vibration) energies. Biphenyl-4-thiol (BPT) is chosen (inset Fig. 1A) since it provides vibrations that are simultaneously active in both IR absorption and Raman, and binds strongly and consistently to gold. Integrated into a dual-wavelength gold antenna, the nanoparticle-on-resonator (NPoR), this strongly confines visible and MIR within the same active region (*19*), accessing single-molecule optomechanical nonlinearities (*20, 21*). The Au disk resonators (diameter 6µm) have a fundamental resonant mode around $\lambda$=10µm and high order modes in the visible (*19*). Onto these is self-assembled a molecular monolayer of BPT, with 60nm Au nanoparticles drop-cast on top. The molecule length sets the 1.3nm spacing (*18*), giving resonances which are experimentally measured with visible and MIR light (Fig. 1C). Comparing with simulations shows field enhancements $E/E_0$ >500 (visible) and >200 (MIR) (Fig. S4A) (*19*), providing a more favourable geometry than previously devised for (simulating) molecular upconversion (*17*). A modified microscope focusses visible and MIR lasers onto the same NPoR device (with >40 NPoRs tested here). The 1080cm$^{-1}$ molecular vibration is observed in SERS antiStokes emission, with amplitude that increases linearly when pumped directly with MIR radiation tuned to the same energy.

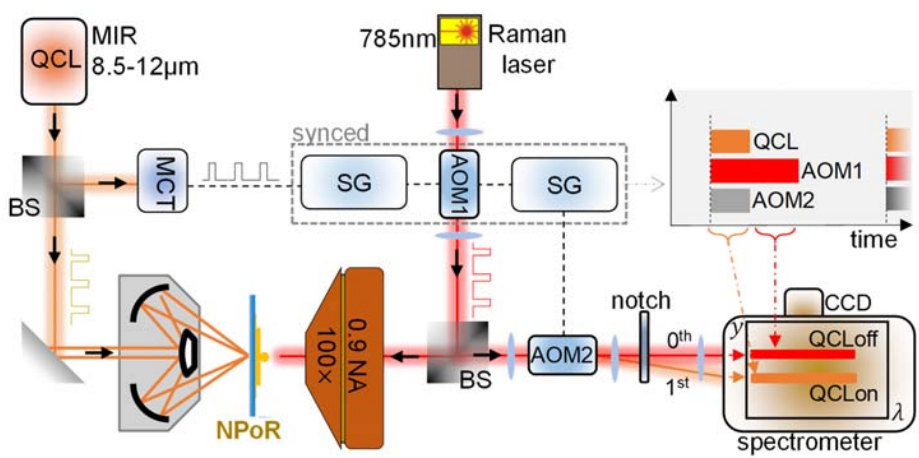

**Fig. 2. MIR and visible spectroscopy.** Dual microscope combines visible probe and MIR pump for frequency upconversion of molecules in nanogaps: AOM (acousto-optical modulator), MCT (mercury-cadmium-telluride detector), BS (beamsplitter), SG (signal generator). Inset: Timing sequence of each repetition of QCL (pump) to Raman laser modulation (AOM1), where AOM2 deflects SERS spectrum to different vertical position $y$ on spectrometer slit.

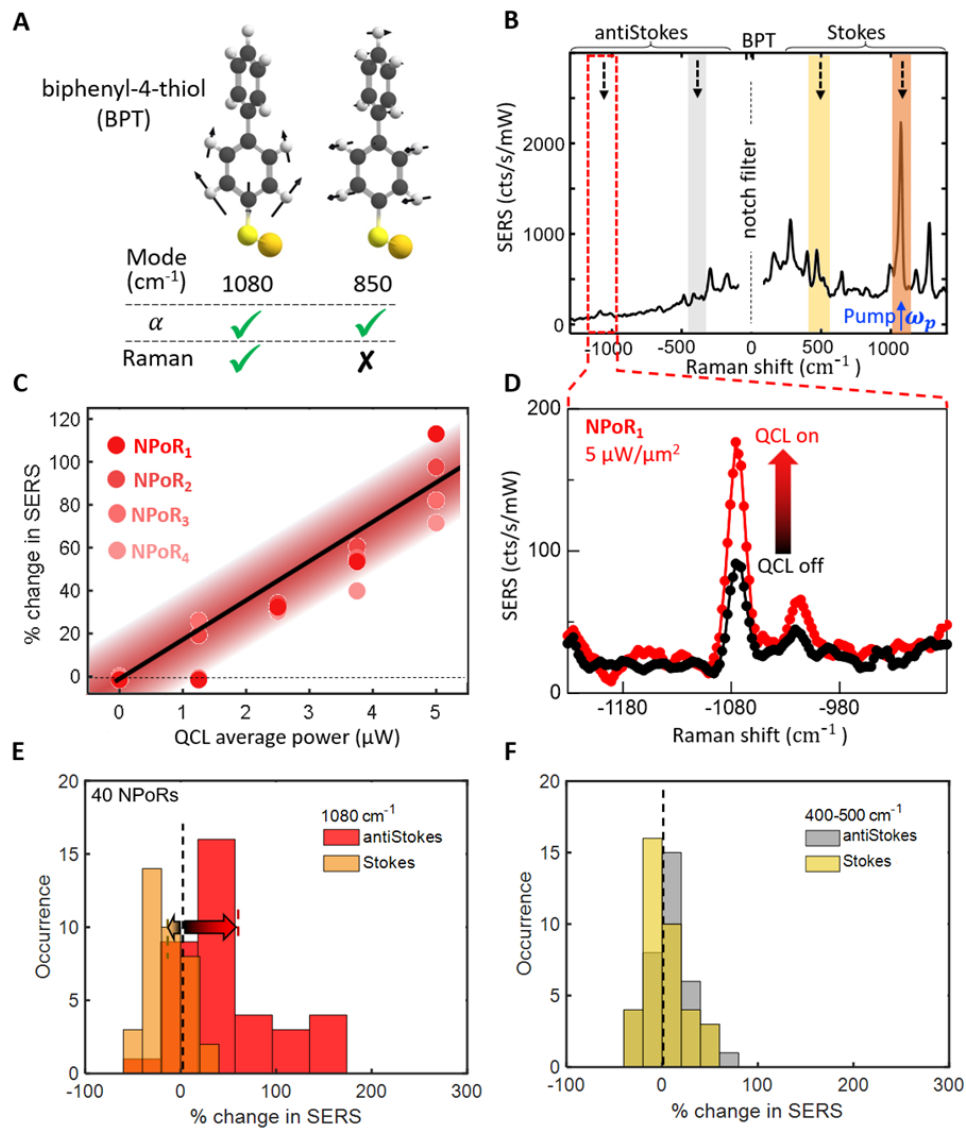

**Fig. 3. Upconversion of MIR to visible photons in hybrid plasmonic antennas.** (**A**) Vibrations of BPT, showing frequencies with strong infrared absorption ($\alpha$) or Raman. (**B**) BPT SERS spectrum from 785 nm probe alone. Shaded regions mark pump (orange, 1080cm$^{-1}$) and monitored frequency bands (arrows). (**C**) Power dependence for 4 NPoRs. (**D**) Raw spectra showing $\nu$=1080cm$^{-1}$ antiStokes increase when MIR pump is on (red). (**E-F**) MIR-induced change in SERS of 40 NPoRs at Stokes and antiStokes peaks at (E) 1080cm$^{-1}$ and (F) 400-500cm$^{-1}$.

Our experiments use synchronised visible and quantum cascade laser (QCL) rectangular pulses (0.4 μs) to collect SERS spectra with/out the MIR light (Fig. 2). These confirm the prediction of upconversion (*17*), using the $\nu$=1080cm$^{-1}$ BPT mode which is both infrared and Raman active (Fig. 3A, 4A). Measuring SERS from NPoRs shows the expected BPT vibrations on both Stokes and antiStokes sides of the laser (Fig. 3B), which are stable and repeatable over long periods. The QCL is then tuned to the same photon energy $h\nu$ (orange, Fig. 3B) and an infrared pump power dependence recorded (Fig. 3C). We find that the antiStokes SERS is 200% higher when QCL average powers of 5μW/μm$^2$ are incident (Fig. 3D, using peak area ratio AS(QCL$_{on}$)/AS(QCL$_{off}$) with background-subtracted antiStokes peaks, see Methods, S2). The expected linear dependence of frequency upconversion with pump power is similar for different NPoRs (red points, Fig. 3C). The lowest detectable light intensity of these dual-wavelength plasmonic antennas is ~1μW/μm² (Fig. 3C), while the lock-in detection synchronised technique here shows that the response speed is sub-μs, much faster than the QCL pulse repetition rate (5 μs).

To better quantify the upconversion efficiency, we measure the percentage change of SERS on 40 NPoRs, where each NP is located at different positions on its disk antenna. These show an average 52% increase of antiStokes at $\nu$ (red, Fig. 3E) for 5μW/μm² MIR, while by contrast the Stokes at $\nu$ shows a decrease of 13% (red). No systematic correlation with the NP position on the disk is apparent, though likely it controls in-coupling of both visible and MIR light into the nanogap.

To confirm the frequency upconversion mechanism, the % SERS changes are also extracted for the 400-500cm$^{-1}$ spectral region (yellow/grey shaded areas for Stokes/antiStokes, Fig. 3B). These low frequency vibrational modes show no discernible change within the ±10% signal noise (Fig. 3F). This lack of low wavenumber signal shows that the signal is not simply thermal heating (Fig. S5), as also suggested by the sub-μs response, but instead is a non-equilibrium response. If simple heating were involved, a trebling of antiStokes at 1080cm$^{-1}$ would give a 60% increase at 450cm$^{-1}$, which is not observed.

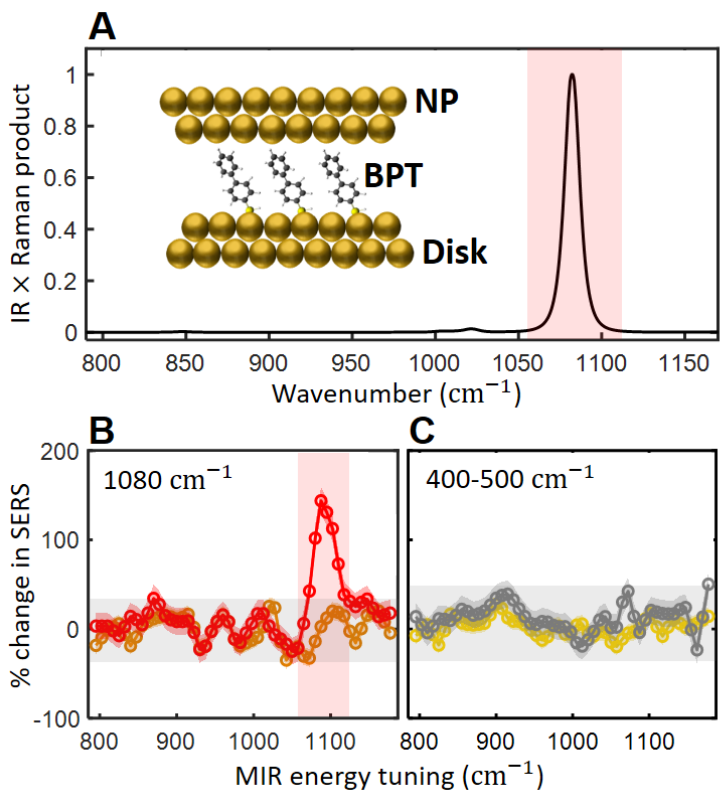

**Fig. 4. MIR tuning dependence of upconversion in NPoR plasmonic construct.** (A) Calculated product of molecular infrared absorption and Raman cross-section for BPT in plasmonic nanogaps (inset). (B-

C) Percentage change in SERS from illuminated NPoR *vs* MIR frequency. Orange (yellow) and red (grey) correspond to SERS change of (**B**) 1080cm$^{-1}$ and (**C**) 400-500cm$^{-1}$ Stokes and antiStokes peaks.

To understand the frequency-selective dependence, we calculate the product of infrared absorption and Raman intensity of BPT, averaged over all orientations for each normal mode (Fig. 4A, SI section S1) (*22*). This clearly shows that the optimum overlap of optical and vibrational modes is at 1080cm$^{-1}$, and that the dipoles are all well-aligned with the vertical $E$ field in the nanogap at both visible and MIR wavelengths. To confirm this, we tune the QCL from 795-1170cm$^{-1}$ in 15cm$^{-1}$ steps, ensuring a constant 5μW/μm$^2$ incident on the sample. While the NPoR device shows a resonant antiStokes increase of 140% at 1080cm$^{-1}$ it shows no increase elsewhere across the frequency scan (red, Fig. 4B), and neither does the Stokes signal (orange). No clear change in SERS intensity is also seen for the 400-500cm$^{-1}$ lines across this MIR tuning on the same NPoR (Fig. 4C). This data clearly distinguishes the direct resonant pumping of the optimum 1080cm$^{-1}$ mode.

The quantum efficiency of these devices is estimated by calibrating to the thermal scale of antiStokes emission. At room temperature ($T$=300K), with MIR powers of $P$=100μW (intensity $I$ =5μW/μm$^2$) at 1080cm$^{-1}$ ($h\nu$=0.13eV), and assuming decay times from the first vibrational state of $\tau$=1 ps (*20, 23*), using the measured antiStokes increase of $\Delta\zeta$ =100% (see Fig. 3C) gives the fraction of MIR photons arriving that produce an observed vibration excitation as

$$\eta = \Delta\zeta \exp\left\{-\frac{h\nu}{k_B T}\right\} \left[\left(\frac{P}{h\nu}\right)\tau\right]^{-1} \tag{1}$$

corresponding to photon quantum efficiency $\eta \sim 2\times 10^{-6}$ in this first generation of devices. The induced occupation of the first vibrational level is estimated to be $\Delta\zeta \exp\{-h\nu/k_B T\} \sim 1\%$. This should be compared with the theoretical estimate (*17*),

$$\eta = \eta_{\text{IR}} \frac{g^2 \tau}{\kappa_{\text{IR}}} \sim 1 \times 10^{-6} \tag{2}$$

using the measured MIR linewidth (Fig. 1C) to get the antenna loss rate $\kappa_{\text{IR}} \sim 2.7$THz, an antenna efficiency $\eta_{\text{IR}} \sim 0.5$ and an optomechanical coupling $g \sim 2$GHz for BPT molecules in the nanocavity gap (*20, 21*). This assumes that the optical cross-section of the dual-wavelength antenna matches the incident MIR focus. The main inefficiency is in the fraction of MIR photons giving significant field inside the NPoR gap to vibrationally excite a molecule. Improving the Q factor of the antenna, for example using hybridization with photonic cavities, is needed for further enhancements (*24*).

We also show it is possible to fabricate these integrated NPoR detectors using SiN waveguides on standard 4" Si wafers (*25*), by cheap and scalable combination of top-down and bottom-up lithographies (Fig. S3). Prospects for wideband operation are promising, with lower frequency antiStokes emission already observed at 250cm$^{-1}$ (see Fig. 3B, $\lambda$=40μm or 7.5 THz). Utilising alternative molecules embedded in NPoRs, SERS lines observed ~160cm$^{-1}$ access targets for astronomical detectors (OH line at 4.7THz and lower). Although the lifetime of such devices is not yet fully characterised, it already exceeds 1 week, depending on suitable encapsulation to exclude oxygen. The rapid relaxation of non-resonant molecules in the virtual Raman process is encouraging for engineering robust performance. We emphasise further increases in sensitivity can come from exploiting single-atom picocavities, which deliver hundred-fold larger SERS signals from the enhanced light localisation around single Au adatoms (*26, 27*), with simple estimates based on eqn (2) using the measured $g \sim$5THz (*20, 21*) giving near unity upconversion efficiencies. This makes current efforts to stabilise picocavities significant, as well as optimising the overlap of MIR light in the same volume.

**ACKNOWLEDGMENTS**

**Funding:** We acknowledge support from European Research Council (ERC) under Horizon 2020 research and innovation programme THOR (829067), POSEIDON (861950) and PICOFORCE (883703). We acknowledge funding from the EPSRC (Cambridge NanoDTC EP/L015978/1, EP/L027151/1, EP/S022953/1, EP/P029426/1, and EP/R020965/1). X.Z. acknowledges support from KU Leuven Internal Funds C14/19/083, IDN/20/014, KA/20/019 and FWO G090017N. R.C. acknowledges support from Trinity College, University of Cambridge. Z.K.B. and E.R. acknowledge funding from EPSRC (EP/R013012/1, EP/L027151/1) and ERC project 757850 BioNet. We are grateful to the UK Materials and Molecular Modelling Hub for computational resources, which is partially funded by EPSRC (EP/P020194/1). **Author contributions:** AX fabricated the devices, performed the experiments and data analysis. ZX obtained analytical electromagnetic calculations. RC and AX did full-wave simulations. SKB performed DFT calculations. EM and AX performed SERS of NPoR on Si chips. ER, AM, JJB designed and supervised the work. All authors discussed the results, provided feedback and contributed to the writing of the manuscript. **Competing interests:** The authors declare no competing financial interest. **Data and materials availability:** All data needed to evaluate the conclusions in the study are present in the main text or the supplementary materials. Source data can be found at:

**SUPPLEMENTARY MATERIALS**

Publications website at DOI:

Materials and Methods

Supplementary Text

Figs. S1 to S7